\documentclass[prl,twocolumn,amsmath]{revtex4}

\usepackage{graphicx}
\usepackage{dcolumn}
\usepackage{bm}
\usepackage{upgreek}
\begin{document}


\title{Oscillatory Energy Exchange Between Waves Coupled by a Dynamic Artificial Crystal}

\author{A. D. Karenowska}

\email{a.karenowska@physics.ox.ac.uk}
\affiliation{Department of Physics, Clarendon Laboratory, University of Oxford, OX1 3PU Oxford, United Kingdom}

\author{V. S. Tiberkevich}
\affiliation{Department of Physics, Oakland University, Rochester, MI 48309, USA}

\author{A. V. Chumak}
\affiliation{Fachbereich Physik and Forschungszentrum OPTIMAS, Technische
Universit\"at Kaiserslautern, 67663 Kaiserslautern, Germany}

\author{A. A. Serga}
\affiliation{Fachbereich Physik and Forschungszentrum OPTIMAS, Technische
Universit\"at Kaiserslautern, 67663 Kaiserslautern, Germany}

\author{J. F. Gregg}
\affiliation{Department of Physics, Clarendon Laboratory, University of Oxford, OX1 3PU Oxford, United Kingdom}

\author{A. N. Slavin}
\affiliation{Department of Physics, Oakland University, Rochester, MI 48309, USA}

\author{B. Hillebrands}
\affiliation{Fachbereich Physik and Forschungszentrum OPTIMAS, Technische
Universit\"at Kaiserslautern, 67663 Kaiserslautern, Germany}

\begin{abstract}
We describe a general mechanism of controllable energy exchange between waves propagating in a dynamic artificial crystal. We show that if a spatial periodicity is temporarily imposed on the transmission properties of a wave-carrying medium whilst a wave is inside, this wave is coupled to a secondary counter-propagating wave and energy oscillates between the two. The oscillation frequency is determined by the width of the spectral band gap created by the periodicity and the frequency difference between the coupled waves. The effect is demonstrated with spin waves in a dynamic magnonic crystal.
\end{abstract}



\maketitle

An artificial crystal is a wave transmission medium which features an ``artificial lattice'' created by a wavelength-scale modulation in its material properties. Such systems---which include, for example, optical photonic crystals~\cite{1,2,3,4,5,6,7} and magnetic magnonic crystals~\cite{8,9,10,11,12,13,14,15}---belong to the class of so-called metamaterials: synthetic media with properties derived from an engineered mesoscopic structure. The wave transmission spectra of artificial crystals typically include band gaps. These gaps arise as a result of resonant wave-lattice interactions analogous to the atomic-scale Bragg scattering phenomena observed in natural crystals~\cite{16}. Originally, the study of artificial crystals was directed exclusively toward time-invariant structures.  However, considerable attention is now focussed on an emerging class of \emph{dynamic} artificial crystals~\cite{1,2,3,4,5,8,13,14} which have properties that can be modified \emph{whilst a wave propagates through them}.

\begin{figure}
\includegraphics{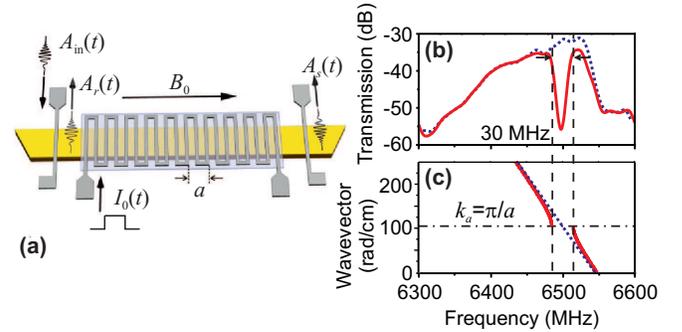}
\caption{(Color online) (a) The dynamic magnonic crystal (DMC) comprises a metallic meander structure with 20 periods of lattice constant $a$ (10 shown) situated close to the surface of an Yttrium Iron Garnet spin-wave waveguide. Microstrip input (left) and output (right) antennas are arranged on the surface of the film and a bias magnetic field $B_0$ is applied along the spin-wave propagation direction. (b) Spin-wave transmission characteristics with the DMC in the ``off'' state (no current applied to the meander structure, dotted, blue) and in the ``on'' state (static current of 1~A in the meander structure, solid, red). In the latter case there is a band gap corresponding to the wavevector $k_a = \pi/a$ centered on $\omega_a=2\pi\cdot6500$~MHz of $-3$~dB width approximately $30$~MHz. (c) Calculated spin-wave dispersion relationship for the DMC system in the ``off'' (dotted, blue) and ``on'' (solid, red) states.}
\end{figure}

A particularly interesting subset of dynamic artificial crystals have a lattice which can be turned ``on'' and ``off''. Recently, it has been established that if such a crystal undergoes a rapid transition from ``off'' to ``on'' whilst a wave mode lying within the band gap is excited \emph{inside} it, this mode becomes coupled to a secondary counter-propagating wave, generally of a different frequency~\cite{14}. In this Letter, we demonstrate that the underlying physics of this coupling mechanism reveals a fundamental result of general wave dynamics. It is shown that the process is one of \emph{oscillatory} inter-modal energy exchange. We explore the features of the effect experimentally using spin waves (magnetic excitations, the quanta of which are known as magnons~\cite{15,17}) in a dynamic magnonic crystal (DMC)~\cite{8,14,15}, and show that results are consistent with a theoretical description of the coupling phenomenon which models the waves as interacting harmonic oscillators.

Figure 1(a) shows the experimental spin-wave system. The one-dimensional DMC comprises a planar metallic meander structure (period $a=300$~$\upmu$m) situated close to the surface of an Yttrium Iron Garnet (YIG) thin-film spin-wave waveguide~\cite{8,15}. An antenna is provided at each end, and a static bias magnetic field $B_0=180$~mT is applied along the spin-wave propagation direction. In the absence of a current in the meander structure, the transmission characteristics and spin-wave dispersion relationship $\omega(k)$ for the system are as shown by the dotted lines in Figs.~1(b) and 1(c), and the DMC is said to be ``off''. If a current $I_0(t)$ is applied to the meander structure, its Oersted field (amplitude $\Delta B_0 \propto I_0$) spatially modulates the waveguide's magnetic bias. Under these conditions, the DMC is ``on'' and has a band gap corresponding to the wavevector $k_a = \pi/a =105$~rad cm$^{-1}$ which appears as a discontinuity in the spin-wave dispersion relationship centered on the resonance frequency $\omega(k_a)=\omega_a=2\pi\cdot6500$~MHz (solid line, Fig.~1(c))~\cite{17,18}. The band-gap width is determined by $\Delta B_0$ ($\sim 1$~mT in our experiments) and thus the current amplitude.

\begin{figure}
\includegraphics{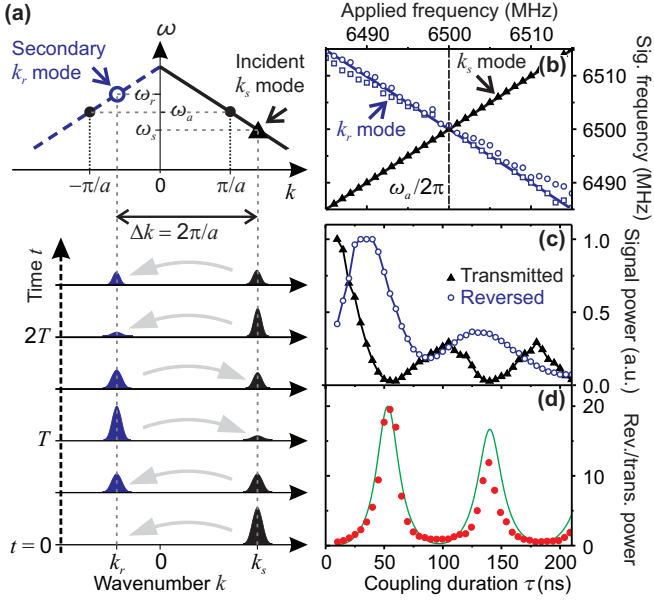} \caption{(Color online) (a)  Inter-modal energy exchange in the DMC system. The diagonal lines in the upper panel represent the spin-wave dispersion curve. Solid dots (black) mark the wavevectors $\pm k_a$ corresponding to the band-gap center frequency $\omega_a$. If the DMC is switched ``on'' at $t=0$ whilst an incident mode $k_s \approx k_a$, frequency $\omega_s$ (black triangle) is inside, this mode is coupled to a secondary mode $k_r = k_s - 2k_a$, frequency $\omega_r$ (blue open circle). In the lower panel, the heights of the peaks centered on $k_s$ (black) and $k_r$ (blue) represent the power in the incident and secondary modes. Almost complete energy exchange occurs every $T$ seconds for as long as the coupling persists. (b)  Reversed ($k_r$, blue open symbols) and transmitted ($k_s$, black triangles) signal frequencies as a function of the applied spin-wave carrier frequency. Reversed signal data is shown for two values of coupling duration: $\tau = 25$~ns (circles) and $\tau = 20$~ns (squares). Transmitted signal datapoints for both values of $\tau$ are represented by a single triangular symbol. Solid lines show the theory of Eq.~(6). (c)  Experimental reversed ($P_r$, open blue circles) and transmitted ($P_s$, black triangles) signal powers as a function of $\tau$ for $\omega_s=\omega_r=\omega_a$.  (d) Ratio of reversed to transmitted signal powers as a function of $\tau$; circular symbols (red) are points corresponding to the experimental data of (b), the solid curve (green) is the theoretical curve (see text).}
\end{figure}

If a spin wave with wavevector $k_s \approx k_a=\pi/a$ is incident on the crystal whilst it is in the ``on'' state, it is simply reflected, creating a ``dip'' in the experimental transmission characteristic (Fig.~1(b)). However, if the crystal is {\em switched} from ``off'' to ``on'' whilst such a wave \emph{is inside}, a geometrically-induced coupling between the incident mode and secondary modes with wavevectors $k_r = k_s\pm2k_a$ is brought about. This coupling is pronounced when the eigenfrequencies of both waves, $\omega_s = \omega(k_s)$ and $\omega_r = \omega(k_r)$, are close to each other, i.e.\ for $k_s \approx k_a$ and thus {\em counter-propagating} waves $k_r = k_s - 2 k_a \approx -k_a$. The dynamics of the complex amplitudes $c_s$ and $c_r$ of the two interacting waves can be modeled by a pair of coupled equations~\cite{19}
\begin{subequations}\label{model}
\begin{eqnarray}\label{eq1a}
	\frac{dc_s}{dt} &=& i \omega_s c_s + i \Omega_a c_r,
\\\label{eq2b}
	\frac{dc_r}{dt} &=& i \omega_r c_r + i \Omega_a c_s.
\end{eqnarray}
\end{subequations}
where $\Omega_a$ is the coupling constant (with units of angular frequency), numerically equal to the half-width of the crystal band gap. Oscillatory solutions to (1a) and (1b) have frequencies
\begin{equation}\label{eq2}
 \omega_{1, 2} = \overline{\omega} \pm \sqrt{\Omega_a^2 + \delta\omega^2} = \overline{\omega} \pm \Omega
\end{equation}
where $\overline{\omega} = (\omega_s + \omega_r)/2$ is the average eigenfrequency of the waves, and $\delta\omega = (\omega_s - \omega_r)/2$ is their frequency mismatch. Equation (\ref{eq2}) describes the dispersion relation for a {\em static} magnonic crystal in the ``on'' state with a band gap of width $2\Omega_a$ centered on the resonance frequency $\omega_s = \omega_r = \omega_a$.

In a {\em dynamic} experiment, at the moment when the crystal is switched ``on'' ($t = 0$) the amplitude of the incident wave is non-zero ($c_s = 1$), whilst the coupled secondary wave is absent ($c_r = 0$). Solving Eqs.~(\ref{model}) with these initial conditions, one can find how the power $P_{s,r} = |c_{s,r}|^2$ of each wave changes with time:
\begin{subequations}\label{powers}
\begin{eqnarray}\label{power-s}
	P_s(t) &=& \cos^2(\Omega t) + \frac{\delta\omega^2}{\Omega^2}\sin^2(\Omega t)
\,,\\\label{power-r}
	P_r(t) &=& \frac{\Omega_a^2}{\Omega^2}\sin^2(\Omega t).
\end{eqnarray}
\end{subequations}
The equations above describe \emph{periodic} energy exchange at frequency $\Omega$ (see Eq.~(\ref{eq2})) between confined, geometrically-coupled incident ($k_s$) and secondary ($k_r$) modes. The maximum power fraction transferred to the $k_r$ mode $R_{\rm max} = \Omega_a^2/\Omega^2 = \Omega_a^2/(\Omega_a^2+\delta\omega^2)$ is of order unity for waves within the band gap ($|\delta\omega| < \Omega_a$) and is achieved after a characteristic energy exchange time
\begin{equation}\label{Teff}
	T = \frac{\pi}{2\Omega} = \frac{\pi}{2\sqrt{\Omega_a^2 + \delta\omega^2}}
\,,\end{equation}
or odd integer multiplies of thereof (see Fig.~2(a)).

If the dynamic crystal is switched ``off'' at $t = \tau$, eliminating the coupling, the waves, which have powers given by the instantaneous solutions of Eqs.~(\ref{powers}) travel in opposite directions. Experimentally, the oscillatory energy exchange process may be studied by recovering the ratio of the powers of these propagating waves,
\begin{equation}\label{ratio}
	\kappa(\tau) = \frac{P_r(\tau)}{P_s(\tau)} = \frac{\Omega_a^2\tan^2(\Omega \tau)}{\Omega^2 + \delta\omega^2\tan^2(\Omega \tau)}
\,,\end{equation}
which is independent (up to a constant factor) of their damping. Note that the coupling effects frequency conversion---transfer of energy from the incident to the secondary mode---through an entirely linear mechanism~\cite{14}. Since this process is efficient only over a narrow range of wavevectors $k_s \approx k_a$, the dispersion relation can be approximated by a first order Taylor expansion, $\omega_s = \omega(k_s) = \omega_a + v_g(k_s - k_a)$, $\omega_r = \omega(k_r) = \omega_a - v_g(k_r + k_a)$, where $v_g = d\omega(k)/dk$ is the group velocity~\cite{20} at $k = k_a$, and we assume a symmetric dispersion law $\omega(k) = \omega(-k)$:

\begin{equation}\label{frequency}
 	\omega_r = 2 \omega_a - \omega_s
\,,\end{equation}
i.e. the detuning of the incident and secondary mode frequencies is {\em inverted} with respect to $\omega_a$.

Experimental investigations into the geometrical mode coupling mechanism were performed using a current pulse $I_0(t)$ of duration $\tau$ to briefly switch the DMC ``on'' whilst an incident spin-wave signal packet of center frequency $\omega_s$ and duration 200~ns excited by a microwave drive signal ($A_{\rm in}$, Fig.~1(a)) was contained within it~\cite{21}. When the DMC is switched ``on'' ($t=0$), transfer of energy from the incident $k_s$ mode into the secondary $k_r$ mode initiates (see Fig.~2(a)), both modes being confined to the crystal. Following the deactivation of the coupling at $t = \tau$, the $k_s$ mode propagates to the output antenna, producing a transmitted signal of power $A_s \propto P_s(\tau)$, whilst the $k_r$ mode travels to the input antenna, originating a ``reversed'' signal of power $A_r \propto P_r(\tau)$ (see Fig.~1(a)). (Note that, in the absence of damping and with perfect antenna efficiency $A_{s, r}=P_{s,r}(\tau)$).

\begin{figure}
\includegraphics{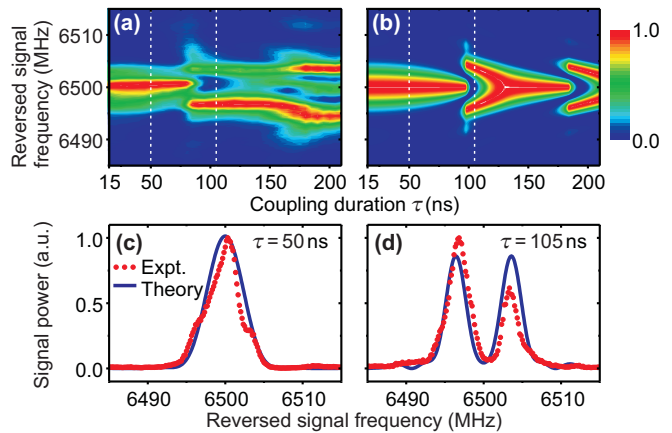}
\caption{(a)  Reversed ($A_r \propto P_r(\tau)$, Fig.~1(a)) signal spectra for varying $\tau$ (input spin-wave signal duration 150~ns, carrier frequency $\omega_s = \omega_a = 2 \pi \cdot 6500$~MHz). Red is indicative of the highest signal intensities, dark blue the lowest, each spectrum is individually normalized. (b) Theoretical calculation of the data of (a) using experimentally derived parameters. Panels (c) and (d) show sections of the experimental data of (a) (red circular symbols) and (b) (blue curves) at $\tau = 50$~ns and $\tau = 105$~ns respectively (vertical dashed lines, (a) and (b)).
}
\end{figure}

Measured frequencies of the transmitted ($k_s$) and reversed ($k_r$) signals are plotted in Fig.~2(b) for two values of the coupling duration $\tau$. The frequency of the transmitted signal is not changed by interaction with the crystal. The reversed frequency, as predicted by Eq.~(\ref{frequency}), is determined solely by the geometry of the DMC and is independent of $\tau$. The data are a near-perfect fit to the theory (solid lines on the same axes), providing strong evidence for its validity. The antiphase variation of $P_s$ and $P_r$ with the coupling time $\tau$, predicted by Eqs.~(\ref{model}), is confirmed by the experimental data of Fig.~2(c) for the resonance case $\omega_s = \omega_r = \omega_a$. Figure 2(d) shows the ratio of the reversed to transmitted signals of Fig.~2(c) (circular symbols) and the corresponding theory ($\kappa = P_r/P_s$, solid line). As can be seen, there is good agreement between the two trends. For the theoretical curve of Fig.~2(d) we assume $\Omega_a = 2 \pi \cdot 5.53$~MHz, a small detuning $\delta \omega = 0.03\%$, and an effective coupling duration $\tau_{\rm eff} = \tau - 2\Delta \tau$ where $2\Delta \tau = 12.5$~ns to account for the effect of the finite rise and fall times of the DMC current pulse. The plot is also scaled to match the experimental and theoretical maximum ratios (absolute values being dependent on losses and antenna efficiencies). The reduction in the signal amplitudes of Fig.~2(c) with increasing $\tau$ occurs as a result of damping. The non-zero minima of the reversed curve results from the finite spectral width of the input signal (which accounts too for the difference in heights of the two peaks in Fig.~2(d), the total power of the reversed signal is the integral of Eq.~(3b) over all frequency components of the input signal) and weak spurious reflection of spin waves from the metal meander structure. In combination with damping, spurious reflections are also responsible for the small offset between the maxima and minima of the transmitted and reversed curves.

Note that the value of the coupling constant $\Omega_a = 2\pi \cdot 5.53$~MHz derived from the dynamic experiments reflects the width of the static DMC band gap at approximately $-20$~dB (Fig.~1(b). This is consistent with the fact that the coupling mechanism requires strong rejection of waves from the crystal.

For a fixed incident mode power, Eqs.~(3) predict that the frequency dependence of the secondary mode power varies with the coupling duration $\tau$. In order to observe this dependence experimentally, the power spectrum of the reversed signal ($A_r(t)$, Fig.~1(a)) produced from an input spin-wave packet of duration 150~ns and carrier frequency $\omega_s = \omega_a = 2 \pi \cdot 6500$~MHz was recovered for varying $\tau$. Results are plotted in Fig.~3(a). Figure 3(b) shows the corresponding theory. As can be seen, the theory accurately predicts the key features of the data, namely the high signal intensity at $\omega_r = \omega_a$ which extends from the shortest measured coupling duration (15~ns) to around $\tau = 90$~ns, and the branching of the spectra at approximately $\tau = 90, 180$~ns. Figures 3(c) and 3(d) are sections through the data of Figs.~3(a) (red circular symbols) and 3(b) (blue curves) at $\tau = 50, 105$~ns respectively. The deviation between theory and experiment at long coupling times in Figs.~3(a) and 3(b) occurs as a result of leakage of energy from the DMC during the coupling interval; an inevitable feature of any experiment using an artificial crystal of finite length. The different peak heights in Fig.~3(d) are caused by a small detuning (0.03~$\%$) of the carrier signal frequency from the resonance frequency $\omega_a$ in combination with a weak frequency dependence of the characteristics of the microstrip antennas.

\begin{figure}
\includegraphics{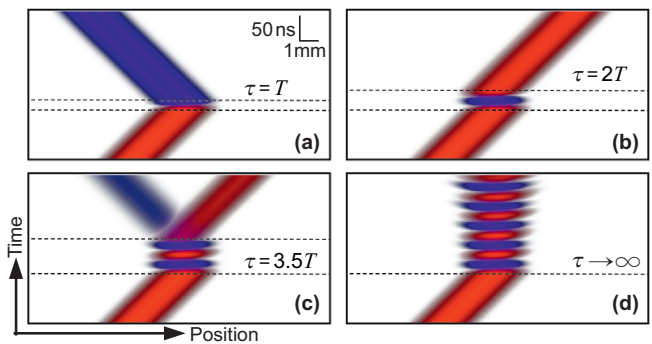}
\caption{Theoretical spatio-temporal maps of interacting waves in DMC system for different coupling durations $\tau$ at $\omega_s = \omega_r = \omega_a$. Each panel shows the space- (horizontal) time (vertical) dependence of the wave power. Red and blue colors correspond to the incident ($k_s$) and secondary ($k_r$) waves respectively and the intensity of the color represents the wave power. Dashed horizontal lines indicate the time interval $\tau$ during which the DMC is ``on''. (a)  $\tau = T$. (b)  $\tau = 2T$.  (c) $\tau = 3.5T$ (50:50 beamsplitter). (d)  $\tau \rightarrow \infty$.
}
\end{figure}

To further clarify the features of the mode coupling mechanism, Fig.~4 shows theoretical spatio-temporal maps of the interacting waves in the artificial crystal system for different coupling durations $\tau$. In each panel, the power of the incident ($k_s$, red) and secondary ($k_r$, blue) modes are plotted as a function of time (vertical) and position (horizontal) and the dashed horizontal lines indicate the time interval $\tau$. An ideal, lossless system with $\omega_s = \omega_r = \omega_a$ is assumed. In the limit $\tau \rightarrow \infty$ the incident wave energy is effectively ``stopped''; it remains confined to the crystal and repeatedly oscillates between the $k_s$ and $k_r$ modes (panel (d)).

In summary, we have described a mode coupling phenomenon in a dynamic artificial crystal. The mechanism, which is a principle of application to waves of any nature, involves \emph{oscillatory} energy exchange between two counter-propagating wave modes, generally of different frequencies. The repeated spontaneous reversal of energy flow direction which characterizes the effect distinguishes it from all other known methods of inter-modal energy transfer in wave systems (see for example~\cite{15,17,22}). Experimental measurements performed using a dynamic magnonic crystal are in agreement with a theoretical description of the mechanism which models the coupled waves as interacting harmonic oscillators. The relationship between the frequencies of the coupled modes is determined by the crystal geometry, and the rate of energy exchange between them by the width of the crystal band gap and their frequency mismatch. As well as representing a fundamental development in the understanding of wave dynamics in metamaterial systems, the effect potentially finds widespread technological application in the context of signal processing.

Financial support from the Deutsche Forschungsgemeinschaft (Grant No.\ SE 1771/2), from the ECCS and DMR Programs of the USA National Science Foundation, and from the U.S. Army TARDEC, RDECOM is gratefully acknowledged. Dynamic magnonic crystals were fabricated with the technical assistance of the Nano+Bio Center, TU Kaiserslautern. A.D.K. is grateful for the support of Magdalen College, Oxford.

\medskip

\noindent Submitted, November 2010.


\begin{thebibliography}{sotief}
\bibitem{18} The orientation of the magnetic bias field (Fig.~1(a)) supports the propagation of backward volume magnetostatic spin waves (BVMSW) in the YIG. As shown in Fig.~1(c), the BVMSW dispersion relationship $\omega(k)$ has a negative rather than a (usual) positive slope. However, we stress that this detail has no bearing on the main result of this Letter which is valid for any symmetric dispersion function $\omega(k)= \omega(-k)$.
\bibitem{21} So long as the spectral extent of the spin-wave packet is restricted to the band gap of the magnonic crystal in the ``on'' state and its spatial extent allows it to be confined within the DMC structure, the results discussed are invariant with the spin-wave packet duration.
\end{thebibliography}
\end{document}